# Decision Towards Green Careers and Sustainable Development

Adam Sulich [a,*], Małgorzata Rutkowska [b], Uma Shankar Singh [c]

[a] *Wroclaw University of Economics and Business, Faculty of Management, Komandorska 118-120, Wroclaw 53-345, Poland*
[b] *Wroclaw University of Science and Technology, Faculty of Computer Science and Management, Wyspianskiego 27, Wrocław 50-370, Poland*
[c] *University of Szczcein, Faculty of Economics, Finance and Management, ul. A. Mickiewicza 64, Szczecin 71-101, Poland*

**Abstract**

The graduates' careers are the most spectacular and visible outcome of excellent university education. This is also important for the university performance assessment, when its graduates can easily find job on the labour market. The information about graduates' matching their qualifications and fields of studies versus undertaken employment, create an important set of data for future students and employers to be analysed. Additionally, there is business environment pressure to transform work places and whole organizations towards more green and sustainable form. Green Jobs (GJ) are the element of the whole economy transformation. This change is based on the green qualifications and green careers which translate theoretical assumptions into business language. Therefore, the choice of future career path is based specified criteria, which were examined by surveys performed among graduates by the career office at Wroclaw University of Technology (WUT) in Poland. The aim of this article was to address the question about most significant criteria of green career paths among graduates of WUT in 2019. The special attention was paid for the GJ understood as green careers. In this article the multi-criteria Bellinger's method was explained, presented and then used to analyse chosen factors of choice graduates' career paths and then compared with Gale-Shapley algorithm results in a comparative analysis. The future research can develop a graduate profile willing to be employed in GJ.

* Corresponding author.
  E-mail address: adam.sulich@ue.wroc.pl

**1. Introduction**

This article is a continuation of the research series carried out among students and graduates of Wroclaw University of Technology (WUT) in Poland which aim was dedicated to recognize gaps in their qualifications and to assess the process of their career paths. In this study Authors decided to replicate the method used in published study of *Malara, Z., Miśko, R., and Sulich, A. (2016) Wroclaw University of Technology graduates' career paths. Vesnik Grodzenskaga Dzâržanaǔga Unìversìtèta Ìmâ Ânkì Kupaly. Ser. 5, Èkanomìka, Sacyâlogìâ, Biâlogìâ. 6 (3), 6–12*. Based on this paper [1] findings, methods and results in this study Authors decided to focus on the decisions towards green careers and Sustainable Development (SD). This paper brings a novelty into scientific discussion, due to the new approach to discuss GJ (or green careers) and the discussion of the decision-making process in this context.

The term "green" and "sustainability" are used usually as synonyms for environmental concern [2]. As emphasized by experts on the subject: "the essence of the SD does not lie in balancing relations between such domains (orders) as economy, society, space or nature, but in choosing the degree of durability, the needs of the economy and society" [3]. However, these terms lack the wider scope in terms of understanding needed for the implementation beyond science careers. These terms are used interchangeably to project their image on the labour market [4,5]. It may help organizations to become more efficient, increase profits and customer base and ultimately contribute towards the well-being of society [6,7]. The specific group of candidates are young educated people, graduates from universities who are one of the high preferred group on segmented labour market [1]. Recruitment teams are interested in this group of candidates because they possess demanded high qualifications and their salary expectations are not so high like among more experienced employees with the same education background [8,9]. In this article the choice made by universities' graduates is most important and is a start point to understand better their decision-making process [1,10].

The goal of this article is to address the question which criteria of career path choice are most important for graduates of WUT for the GJ (or green careers). Second goal is to present and use so modified multi-criteria Bellinger's method to analyse chosen factors of career path choice among graduates of WUT [1,11]. Results obtained from calculations in presented Bellinger's method were again compared to results obtained from Gale-Shapley algorithm to check is there any difference in matching between proposed job positions (p) and criteria of choice (c) and correctness of chosen method. The reference point were the findings in the Malara et. al. cited paper [1].

The text is organised as follows: The first part describes and recalls few important for decision process game theories and their potential scope in career path choice along with the SD context. The following section distinguishes role of choice factors and among them some related to the SD indicators are mentioned. The next part presents the Bellinger's method and analyses data obtained from research conducted in 2015 and 2019. The results section with the comparative analysis of two surveys (performed in 2015 and 2019) is presented. Readers who are more interested may find a broader survey of the subject in another works [1,12]. The final section contains brief conclusions and question which should be addressed in further studies.

**1. Literature Review**

On the labour market at least two main and active participant's sides can be observed. These labour market attendants vividly make their decisions [1]. One of them are organizations, which are looking for best university graduates in the recruitment process. Second are candidates looking for their first job or future candidates which decide to which employer they should apply when changing their workplace [6]. Both these groups (employers and employees) make their decision based on group intertwined and mostly subjective criteria. However, they are looking for best and most close part to become their "perfect candidate" or the "ideal employer" respectively [1,13].

Career choices are individual preferences lead by many factors, where the personal acceptance is the most important for sustainability or continuity in the job with satisfaction [1]. Year after year, graduating students are entering the job market to build their professional career with an open mind-set, however affected by some trends. Then global population is growing at a faster pace and is demanding rights to jobs and pushing governments for policy restructuring in both public as well as private spheres [14]. The contemporary decade is witnessing many challenges, but green career choices are growing at a faster pace not only if the strategy or the business model is discussed but also when it comes to the decision-making or management style [15,16]. The shift towards the green career is the need for the SD, promoting the quality and quantity of work with a higher satisfaction [17]. Green employment generation can be possible only with social inclusiveness and environmental sustainability as the core for all activities with the global contribution towards career [18,19].

Green career selection is seen nowadays as a proactive approach to cope up with upcoming challenges during the transition period which explores the opportunities and become a participant in change [20]. International Labour Organization (ILO) is continually emphasizing on strengthening the policy for interconnection between efficient strategies and green employment for sustainability while stimulating green initiatives as the catalyst [21,22]. Green careers can be a chance for youth to participate in job market with a higher sensibility and is one of the ways to contribute towards the well-being of the society [23]. Government agencies must plan for short- and long-term implementation of measures considering student's investment in education, individual choices and labour market requirements for a successful model of green employment [3,24]. The opportunities can be explored by formulating policies such as stimulating future employers to consider green careers as compared to traditional careers, counselling the youth to opt the green career [25,26]. The Green career choices may face many challenges because of fresh graduate's expectations from job and company and being less lucrative and competitive, lack of linkage in educational learning and job requirements, and slow career growth with invisible future goals [27,28]. The reform in the economy and policy for enterprises pinpointing the concept of green career can bridge the gap between the expectation of the employer and the graduate applicants in a sustainable or green economy [4].

The European Union has shown a proactive approach in providing support to the green economy [6,29]. The requirement for green careers is considered as the key to the SD [1,30]. The global challenges can get tackled only by aligning the new generation towards understanding the importance of green careers [31,32]. The SD achievement is not a one-day task and to reach it, there must be a very systematic plan with the inclusiveness of the global population [6,22]. There must be connectivity between the market and production for the consumption of green products and services [33]. The major hurdle for the selection of green careers is the consumption market [34]. If the market accepts the product or services, students will be more motivated towards joining the green careers [14]. Academic orientations are also changing worldwide to create a competent workforce to participate in green careers. Today, sustainability and green career options have been added to the academic curriculum [35]. Students are prepared academically to cope up with market demand in green careers [20]. The core of academic courses is the inclusion of green orientation like green marketing, green finance, and green strategy, transforming the academic environment to align with the transforming economy globally [28,36]. Business entities are using also green image to make it more lucrative and attractive to customers. It is a way to create awareness among people in the market about the benefits of getting green [37]. However, the green marketing and based on this image idea is still in the process of development where the evaluation criteria and contributions are not very clear [13,38].

The decision-making process is generally a single-handed execution but, in most cases, backed by many circumstances and people's opinions [39]. It is applicable in all kinds of decisions either intellectually or practically. It is a strategic process that encompasses many steps [40]. The human process of decision-making is a slow process and time-consuming too. Many studies have shown several techniques for decision-making to solve the problems of conflicting criteria [5,41]. Decision-making models are used purposefully as per the requirement, though all decision-making models are not fully equipped. Each one has its advantages and disadvantages. Decision-making is nothing but choosing the best alternative among the available options [42]. The comparison of alternatives for choosing the best option is executed rationally at the cognitive level. The current study also is based upon decision-making choices for green career selection. Decision-making is a problem-solving approach that can be seen differently in different cases, such as, for a manager regarding managing an organization for organizational culture and future, and for a student or individual the same for choosing a career [18,43]. All decisions are based on some underlying criteria shaped based on the information related to the projected outcome [15,44].

The decision about the choice of career path is one of the element game theory explored mostly by John Nash, Thomas Ferguson, David Gale and Lloyd Shapley [8,11]. In 1951 J. Nash explained that specific processes, such like choice of company or choice of employer, can be described by equilibrium, which is not Pareto optimal [1,15]. Nash's equilibrium states that preferences are most important factor of these decisions (in zero sum game, which mean that someone have to stay unemployed, when others get their jobs). Gale and Shapley described their own stable matching theory however they developed an algorithm leading to equilibrium state (in nonzero sum game) [45]. This theory assumes that each decision leads to stable relation between candidate and employer (employment). The algorithm to reach matching is finite, which is important because both numbers, number of candidates and companies (opened positions), are irrelevant. Therefore, candidates are always on better position, according to these two theories, because they always have right to choose the best for them employer (or even establish their own business). On the other hand, companies cannot make choice of the all-possible candidates from job market, but only among these who applied for specified position. Recruitment is a nowadays more marketing part of employment, because recruiters want to attract as many candidates to enable rational selection which is computed, and managed by computers. Companies' choice is limited then by number of candidates, who wanted ever apply for opened position [11,46].

Industries are aware about government regulations and the global slogan to go green. Students graduating from universities are also inclined towards green careers. The BLS report says that the green job market will grow as much as 126% in different sectors by 2026 for all US jobs, which is higher than the average growth rate of 6.5%. Today environment and nature has been the most important concern for all business entities and is considered as the backbone for sustainability [23]. The enterprises' needs to be socially, economically and environmentally sustainable and the integration of these three can generate green career opportunities. Industries are also focusing on green positions to create opportunities for them in the market and opportunities for new graduates in their companies [47].

Research conducted among students and graduates of WUT in 2015 (N=1242) showed that there is no one main factor of choice and they are all intertwined, related to each other [1,11]. However young people expect every year high salary towards employers, not only in financial matter. Young graduates of WUT have high requirements towards their jobs, but do not revere employment and are not prone to make it to the top of the career ladder irrespective of the price. Still the amount of time and energy consumed in order to cater for professional development might lead to different conclusions. The list of factors of choice future career path was ranked by students who choose eleven most important criteria from over twenty listed in poll form. This result allowed to analyse them with two methods: Bellinger's method and Gale-Shapley algorithm, because they use selected previously criteria to analyse best possible choice [1]. The multi-criteria method was founded by B. Bellinger [48,49], but in Poland developed and propagated actually at the WUT [50–52]. B. Bellinger was using this tool in West Germany Banks in 1970's [48]. He ranked list of customers (applying for a loan) in order from the most reliable to the least reliable one.

It is remarkable in further analysis of the same student group that graduates' career plans and criteria hierarchy are subject of change. Graduates often change in time their criteria, or add new which were previously rejected or neglected (like work life balance, job stability, lack of anxiety and stress, large degree of independence, long holiday leave, prestigious position). Fresh graduates of WUT usually do not start family life early, so some values (time for family, or benefits dedicated for home) become more important in time. Therefore in 2019 (N=1045) research was repeated and replicated among the WUT students and graduates.

In this article decision process is nonzero sum game, which leads to choose ideal job position by graduates and some of these jobs are green careers. The choice is based on many criteria, and each of candidate ranks his own creating a list. Therefore, multi-criteria method here can be used, to check which of these criteria are most important to whole group [1].

## 2. Materials and Methods

Bellinger's method allows to compare evaluation results where different criteria were used, mainly it is useful because of their different labels [51] . It consists of several steps shown below [1,8]:
1. criteria which will be used for decision making are chosen,
2. identifying the measuring units and its desirable changes (in given criteria),
3. determine the lower and upper limit change of each criterion,
4. subjective choice of decision maker, who gives meaning to criteria by specifying weights, so that the sum of all weights is equal to one,
5. creation of a table containing the actual values of the criteria for all variants
6. present any number from table formed in step 5 as a percentage of the "way" from the least desirable to the most desirable state. It should be started by determining the size of the entire "path" from the least desirable to most desirable state for given criteria. So, differences between the states are calculated. Then, from the actual value of the criteria we subtract the value of the least desirable state (calculating actually covered "way"). Finally, we divide actually covered "way" by entire "path".
7. number obtained in step 6 is multiplied by the weight from the step 4.
8. determine the best variant by summing ratings from previous step for each variant.

It is assumed that students who are looking for a job are also following some criteria. Because of limits of this paper, it is decided that there are only 5 job positions, hereafter referred as p1, p2, p3, p4, p5. It is not necessary to show details of those positions to use the method. All possible combinations between number of job position (p=5) and criteria which can be chosen (c=11), can be calculated as follows shown as (1):

$$\binom{c}{p} = \frac{c!}{p!(c-p)!} = \frac{11!}{5!(11-5)!} = 462$$

(1)

A Bellinger's method allows to choose only most significant examples, what makes method much simple to analyse and friendly.

In the first step it is necessary to set criteria which will be used to make decision. According to research of Wroclaw University of Technology [8] there are following criteria:
1. possibility of career development
2. earnings (appropriate pay as net salary)
3. job position related with finished studies,
4. skills and interests (green jobs matching skills),
5. sustainable reputation of the company,
6. transparent career paths (possibility of fast promotion)
7. market success of the company,
8. green experience possible to gain in the organization
9. social benefits
10. localisation of the company
11. size of the company

It was decided that criteria mentioned above, will be labelled sequentially from c1 to c11. Second step requires measuring units and direction of change:
- c1 numbers, growth desired
- c2 thousands of polish zloty, growth expected
- c3 numbers, growth desired
- c4 numbers, growth desired
- c5 numbers, growth desired
- c6 numbers, growth desired
- c7 numbers, growth desired
- c8 numbers, growth desired
- c9 thousands of polish zloty, growth expected
- c10 kilometres from home to location of the company, decrease desired
- c11 numbers, growth desired

Next step, where the lower and upper limit for each change criteria was adopted as follows (conventionally and arbitrarily for this study) [1,11]:
- c1 numbers 1-5
- c2 thousands of polish zloty 2502,87-3238,53
- c3 numbers 1-2
- c4 numbers 1-5
- c5 numbers 1-5
- c6 numbers 1-2
- c7 number 1-5
- c8 number 1-5
- c9 thousands of polish zloty 200-800
- c10 km from home to the company 0-762,4
- c11 numbers 1-4

The fourth stage is to determine weights for all criteria. This is an important analysis stage, because it significantly affects the final results. It was decided to rely on the results of the research WUT. It means that the highest weight is assigned to position of criteria.
- c1, 0.107
- c2, 0.105
- c3, 0.101
- c4, 0.10

- c5, 0.095
- c6, 0.092
- c7, 0.088
- c8, 0.086
- c9, 0.085
- c10, 0.083
- c11, 0.059

In a fifth step is connected with formation of the table that contain the actual values of the criteria for each variant as presented in Tables 1, 2 and 3. These tables present the results of calculations.

**Table 1.** Actual of values for criteria

|     | p1      | p2   | p3      | p4    | p5   |
|-----|---------|------|---------|-------|------|
| c1  | 5       | 4    | 3       | 4     | 1    |
| c2  | 2502,87 | 3000 | 3238,53 | 2900  | 2700 |
| c3  | 2       | 1    | 2       | 1     | 1    |
| c4  | 1       | 3    | 4       | 4     | 3    |
| c5  | 2       | 1    | 4       | 5     | 3    |
| c6  | 2       | 1    | 1       | 2     | 1    |
| c7  | 5       | 4    | 1       | 4     | 4    |
| c8  | 5       | 4    | 3       | 4     | 1    |
| c9  | 500     | 400  | 800     | 200   | 300  |
| c10 | 0       | 50   | 125     | 762,4 | 200  |
| c11 | 1       | 4    | 3       | 3     | 3    |

Source: authors' own work

**Table 2**. Numbers form previous step as a percent of a distance

|     | p1   | p2   | p3   | p4   | p5   |
|-----|------|------|------|------|------|
| c1  | 1    | 0,75 | 0,5  | 0,75 | 0    |
| c2  | 0    | 0,67 | 1    | 0,53 | 0,26 |
| c3  | 0,25 | 0    | 0,25 | 0    | 0    |
| c4  | 0    | 0,5  | 0,75 | 0,75 | 0,5  |
| c5  | 0,25 | 0    | 0,75 | 1    | 0,5  |
| c6  | 1    | 0    | 0    | 1    | 0    |
| c7  | 1    | 0,75 | 0    | 0,75 | 0,75 |
| c8  | 1    | 0,75 | 0,5  | 0,75 | 0    |
| c9  | 0,5  | 0,33 | 1    | 0    | 0,16 |
| c10 | 1    | 0,93 | 0,83 | 0    | 0,73 |
| c11 | 0    | 1    | 0,66 | 0,66 | 0,66 |

Source: authors' own work

**Table 3.** Values after consideration of weights.

| weights | criterion | p1    | p2    | p3    | p4    | p5    |
|---------|-----------|-------|-------|-------|-------|-------|
| 0.107   | c1        | 0,107 | 0,080 | 0,053 | 0,080 | 0     |
| 0.105   | c2        | 0     | 0,070 | 0,105 | 0,056 | 0,028 |
| 0.101   | c3        | 0,025 | 0     | 0,025 | 0     | 0     |
| 0.1     | c4        | 0     | 0,05  | 0,075 | 0,075 | 0,05  |
| 0.095   | c5        | 0,023 | 0     | 0,071 | 0,095 | 0,047 |
| 0.092   | c6        | 0,092 | 0     | 0     | 0,092 | 0     |
| 0.088   | c7        | 0,088 | 0,066 | 0     | 0,066 | 0,066 |
| 0.086   | c8        | 0,086 | 0,064 | 0,043 | 0,064 | 0     |
| 0.085   | c9        | 0,042 | 0,028 | 0,085 | 0     | 0,014 |
| 0,083   | c10       | 0,083 | 0,077 | 0,069 | 0     | 0,061 |
| 0,059   | c11       | 0     | 0,059 | 0,039 | 0,039 | 0,039 |

Source: authors' own work

In the last step it is necessary to sum up the individual values for each variant. Thanks that a final evaluation is obtained.

**Table 4.** Total ranking and the best variant.

| Variant | Total rating |
| --- | --- |
| p1 | 54,75 |
| p2 | 49,65 |
| p3 | 56,67 |
| p4 | 56,87 |
| p5 | 30,63 |

Source: authors' own work

## 3. Results and Discussion

A Bellinger's method showed that the best choice (in given criteria) is p4 offer, because it has the highest score from all available job positions. The highest ranked match in Gale-Shapley's algorithm was pair c1-p4. Both results suggest that criterion of possibility to career development is most important regardless of salary criterion (matching c2-p3). Surprisingly education factor (matching c3-p1) is not most important factor, what can be explained by seeking employment in various activities and professions [45].

**Table 5**. Matching results calculated by Gale-Shapley's algorithm with values form Belinger's method ranked.

| Weight of criterion | Criterion | Job position | Total rating |
| --- | --- | --- | --- |
| 0.107 | c1 | p4 | 56,87 |
| 0.105 | c2 | p3 | 56,67 |
| 0.101 | c3 | p1 | 54,75 |
| 0.1 | c4 | p2 | 49,65 |
| 0.095 | c5 | p5 | 30,63 |

Source: authors' own work

Decisions taken towards green careers are characteristic of the implementation of environmentally friendly solutions. In this sense, they are the implementation of the concept of balanced and sustainable development. Development of the so-called Green careers cover many thematic areas, such as, for example, climate change, renewable energy sources, or changing the consumption and production pattern to a more sustainable one. Therefore, decisions taken towards green careers are to contribute to the impact of the green economy on the improvement of the quality of human life and the environment.

Research is based on the decision-making criteria for the selection of Green Career. The multi-criteria decision making is the best-suited method of choosing the best criteria order of many options refused here. The decision-making process has been through eight steps, where each step is crucial for selecting the best outcome. Steps are as criteria selection, measuring units for the chosen criteria, determining the lower and upper limit, specifying weights for criteria, tabulation of criteria with values, sorting of criteria based on calculated values and "path", multiplying the criteria with weights, and the last is determining the best variant.

The current discussion is based on the assumption that young students also follow some specific criteria for their job selection. The research has considered five job positions (p=5) and eleven criteria (c=11). The Bellinger's method has been important for choosing the most significant example. All 11 criteria used to make decisions are from Wroclaw University of Technology [8]. These criteria further labeled as c1 numbers 1-5, c2 thousands of polish zloty 2502,87-3238,53, c3 numbers 1-2, c4 numbers 1-5, c5 numbers 1-5, c6 numbers 1-2, c7 number 1-5, c8 number 1-5, c9 thousands of polish zloty 200-800, c10 km from home to the company 0-762,4, and c11 numbers 1-4 with their limits for lower and upper.

As per the method, weights determined and allocated for all criteria, which are affecting the result. A matrix presented in Table 1, with jobs and criteria with their allotted weights. Based on the calculation, another matrix presented in Table 2 with determined numbers and values. Table 3 is the presentation of jobs and criteria with weights determining the result. A total rating is calculated for each job and presented in Table 4. Based on the Bellinger's method p4 is chosen as the best job position scoring the highest value compared with the other four job positions. Gale-Shapley's algorithm is used for the comparison, where pair c1-p4 has the highest match showing that career development is the most important criterion for young students for their career choice.

## 4. Conclusions

B. Bellinger's method was presented in this article with its details based on the survey among WUT students in 2019. It is useful tool for facilitating multi-criteria decision. Authors evaluated results using different criteria and compare them (even they are measured in different scale). The advantage of this method is its simplicity and ease of use. The procedure does not require complex and time-consuming calculations [1]. It can be automated using spreadsheet, so it seems to be a useful tool. It has also some disadvantages, because partly relies on a subjective assessment, and choice of random examples to analysis. Despite of limitations which Bellinger's multi-criteria method possesses, which are: arbitrarily assigned criteria and their significance, this method can be assessed as useful to analyse multi criteria decision like choice of career path. In future rules of respondents' choice of certain criteria should be more specified and maybe even there should be a more objective list of them created by some independent committee before their choice, to allow answerers on choose weights of ranked criteria.

The graduates' career paths are not really connected with matching chosen job to graduated field of studies. Therefore, young graduates' careers are not always best benchmark for assessment of university performance. Results obtained in this paper proved that majority of students possess big awareness of the significance of self-development, because they selected education path is to ensure future employment. Young graduates of WUT become well-versed in the requirements of the labour market very quickly and formulates their own opinions.

The scientific contributions of this paper are as follows: analysing of the factors of choice future career of graduated Wroclaw University of Technology; presenting the B. Bellinger's method and its advantages and limitations; describing most important factor of choice future career 3. Mathematical verification of decision process.

**Acknowledgements**

The project is financed by the Ministry of Science and Higher Education in Poland under the program "Regional Initiative of Excellence" 2019–2022, project number 015/RID/2018/19, total funding amount 10,721,040.00 PLN.

The project is financed by the National Science Centre in Poland under the programme "Business Ecosystem of the Environmental Goods and Services Sector in Poland" implemented in 2020-2022 project number 2019/33/N/HS4/02957 total funding amount 120,900.00 PLN.